\documentclass[twocolumn,amsmath,amssymb,pra]{revtex4}
\usepackage{epsfig}
\usepackage{color}
\usepackage{graphicx}
\usepackage{bm}
\usepackage{epstopdf}

\begin{document}

\title{Statistical properties of spectra  in harmonically trapped spin-orbit coupled systems}

\author{O. V. Marchukov, A. G. Volosniev, D. V. Fedorov, A. S. Jensen, N. T. Zinner}
\affiliation{Department of Physics and Astronomy, Aarhus University, DK-8000 Aarhus C, Denmark}

\date{\today}

\begin{abstract}
We compute single-particle energy spectra for a one-body hamiltonian
consisting of a two-dimensional deformed harmonic oscillator
potential, the Rashba spin-orbit coupling and the Zeeman term. To
investigate the statistical properties of the obtained spectra as functions of
deformation, spin-orbit and Zeeman strengths we examine the distributions of the
nearest neighbor spacings. We find that the shapes of these distributions 
depend strongly on the three potential parameters. We show that the obtained shapes in some cases can be
well approximated with the standard Poisson, Brody and Wigner
distributions. The Brody and Wigner distributions
characterize irregular motion and help identify quantum
chaotic systems. We present a special choices of deformation and 
spin-orbit strengths without the Zeeman term which provide a 
fair reproduction of the fourth-power repelling Wigner distribution. 
By adding the Zeeman field we can reproduce a Brody distribution, which is 
known to describe a transition between the Poisson and linear Wigner distributions.  
\end{abstract}

\maketitle

\section{Introduction}
\label{intro}

Cold atomic gases are routinely trapped in spatially confining
harmonic oscillator potentials \cite{ketterle2008, bloch2008,
  esslinger2010, cirac2012}.  The deformation of the trap gives us the
opportunity to consider systems in lower dimensions. In this paper we
discuss particles trapped in a two-dimensional (2D) harmonic trap
which can be deformed (unequal trapping frequencies in transverse
directions) and as a limiting case the one-dimensional (1D) harmonic
trap.  In addition, we put a spin-dependent one-body potential
\cite{dalibard2011} that couples spin and motional degrees of freedom
into a spin-orbit coupling. More specifically we discuss the famous
spin-orbit term introduced by Rashba \cite{rashba1984}.  
This type of spin-orbit coupling has not yet been realized with cold
atomic gases, but there are recent proposals on how to proceed
\cite{anderson2012, goldman2013}.  

  In recent years other spin-orbit coupled systems were realized
  in state-of-the-art experiments for both bosonic
  \cite{lin2009a,lin2009b,lin2011,aidelsburger2011,zhang2012} and
  fermionic systems \cite{wang2012,cheuk2012}.  The case of spin-orbit
  coupled particles confined in a spherical harmonic oscillator trap
  have recently also drawn much attention from different groups
  \cite{ghosh2011, sinha2011, liu2012, ramachandhran2012}. In
  particular, large spin-orbit strengths are studied in the spherical
  trap where Landau-like levels then emerge. A more general anisotropic
  spin-orbit coupling in a $2D$ spherical trap has also been studied\cite{stanescu2007, stanescu2008}.
  Variation of the anisotropy changes the hamiltonian from Rashba to Dresselhaus
  coupling \cite{dresselhaus1955} which both result in identical
  spectra.  In between one of the two spin-orbit terms has zero
  strength which implies a shifted $1D$-oscillator spectrum with an
  offset of zero-point energy from the other direction.  Thus, this
  variation is very similar to our variations where the spin-orbit is
  isotropic and the oscillator becomes strongly deformed.
  Sufficiently large deformation and isotropic spin-orbit is then
  similar to spherical oscillator and sufficiently large anisotropy of
  the spin-orbit coupling.  

A two-dimensional spin-orbit coupled system has several
interesting properties when tuning
deformation, spin-orbit strength, and an additional Zeeman field.  
The resulting energy spectra exhibit large
variations as function of the tunable
external parameters \cite{marchukov2013}.  
Specifically, a given hamiltonian produces a
density of states varying from extremely few to many states
as function of energy. Under perturbations coming from interactions, 
low and high density of states can reflect
stability and instability, respectively
\cite{brack1972}. This can be related to the statistical properties of
the single-particle eigenvalue spectrum.

In Ref.~\cite{marchukov2013} a broad variety of spectra appeared
depending on the controllable parameters of the external field.  This
means that the dynamical properties of the system can be flexibly and
accurately tuned and investigated through appropriate analysis.  
Regular motion arising from integrable systems corresponds to a Poisson
distribution for the nearest neighbor spacing (NNS). For the deformed
harmonic oscillator this is not the case and the distribution depends
on frequency ratios \cite{berry1977}. For irregular (chaotic) motion
the eigenenergies avoid crossing appear between level
separations.  Then the celebrated Bohigas-Giannoni-Schmit conjecture
\cite{bohigas1984} states that the statistical behavior of the levels
can be described by random matrix theory.  A semiclassical explanation
can be found in Ref.~\cite{heusler2007}.

The study of chaotic behaviour in cold atoms dates back at least 
two decades when it was suggested to use atoms in time-dependent 
optical lattices \cite{graham1992}. This proposal was realized 
shortly after by the Raizen group \cite{moore1994}. Later on the 
paradigmatic quantum delta-kicked rotor model found an experimental 
realization with atom optics \cite{moore1995} (see \cite{darcy2004} 
for a more recent discussion of the experimental and theoretical 
investigations into this model). Experiments with large Cesium 
alkali atoms have also been performed \cite{klappauf1999,steck2001} and 
used to study localization and its relation to quantum chaotic 
phenomena. Common to most of these experiments is that the 
key quantity that is measured is the momentum distribution of the 
atoms by time-of-flight which is a well-established technique 
in cold atoms \cite{ketterle2008,bloch2008}. The systems we consider
in this paper can also be probed by mapping out the momentum 
distributions. Some more recent proposals to probe chaotic
dynamics involve both cold atoms, molecules and ions \cite{stone2010,grass2013,mumford2014},
and just a few months ago atoms with large magnetic dipole moments have 
shown experimental signs of chaotic behavior in their scattering dynamics \cite{frisch2013}.

The purpose of the present paper is to provide a statistical analysis
of the energy spectrum arising from spin-orbit coupled non-interacting
particles in a trap.  We shall extract the properties covering regular
as well as chaotic motion.  More specifically we shall calculate the
statistical distributions characterizing the spectra depending on the
tunable parameters.  This in turn points at the possible dynamical
behavior. The paper is organized as follows: In Section
\ref{sec:formulation} we first briefly give the connection between
statistical properties and dynamical behavior. Then, after specifying
the parameters of the system, we describe the adopted statistical
treatment.  Our numerical results are presented and discussed in
Section \ref{sec:results}.  Here we show the nearest neighbor spacing
distribution for different values of trap deformation and spin-orbit
coupling and Zeeman strengths.  Finally, Section \ref{sec:conclusions}
contains summary, conclusion and perpectives for future directions of
research in this area.

\section{Ingredients}\label{sec:formulation}
Statistical analyses of spectra are meant to suggest which type of
dynamical behavior is inherent to a given system.  We therefore first
elaborate on qualitative implications and possible distributions and
their origin in connection with the present system.  In the next
subsections we specify hamiltonian, external parameters and describe
the employed statistical procedure.

\subsection{Motivation and perspective}

Dynamical evolution of a non-stationary state for a given hamiltonian
is determined by the time dependent relative phases of the stationary
states involved.  Many unrelated energies quickly lead to loss of
memory of an initial non-stationary state, and the Poincar{\'e} recurrence
  time becomes large tending to infinity when many states contribute.
  Future properties of the system becomes impossible to predict unless
  at least some average properties of the full spectrum is known.
  Vice versa, strongly related (for example equidistantly distributed) sets of
 energies lead to a regular oscillatory behavior, and predictions become entirely
possible.  These observations are directly relevant for one particle
moving in an external one-body potential.  The relevant energy range
is then the width of the wave packet constituting the initial state.  

For $N$ non-interacting identical particles, where each is subject to
the same one-body hamiltonian, the same spectrum is the basic
ingredient.  The relevant energy range is now obtained from the
distribution of single-particle energies in the initial symmetric
(bosons) or antisymmetric (fermions) many-body state.  For fermions in
their ground state the single-particle energy range is centered around
the Fermi energy. When small uncertainties are inherent in the external parameters we
again encounter possible dynamical evolution of an initally defined
wave packet of given energy and width. However, now these parameters
are as well allowed to vary with time, presumably within rather
restricted limits. We do not envisage this type of complicated
dynamics, although our analyses and the results in this paper would
have some bearing on these as well.

The dynamical behavior of a given system is necessarily contained in
the quantum mechanical stationary solutions, and specifically in the energy
spectrum.  It is therefore of interest to investigate the statistical
properties of the energy spectra as function of the parameters in the
problem.  In particular, the nearest neighbor spacing (NNS)
distribution of energies carries information about the
dynamical behavior~\cite{guhr1998}.  
The distributions emerging by following the statistical procedure can
be compared to dimensionless standard distributions obtained from
random matrix spectra with specified symmetries.  We shall compare our
results with the well known distributions named Poisson, $P_P$, and
Wigner distributions where the latter appears in three variations,
$P_{W1}$, $P_{W2}$ and $P_{W4}$.  The variable, $S>0$, is the energy
difference between nearest neighbors in the energy spectrum.  For
numerical comparison we specify these normalized distributions, i.e.
\begin{eqnarray} 
 P_P(S) &=& \mathrm{exp} \left ( -S \right ), \label{poisson}  \\ 
 P_{W1}(S) &=& \frac{\pi S}{2} \mathrm{exp} \left( - \frac{\pi S^2}{4}\right), \label{wigner1} \\ 
 P_{W2}(S) &=& \frac{32 S^2}{\pi^2} \mathrm{exp} \left( - \frac{4 S^2}{\pi}\right), \label{wigner2}\\ 
 P_{W4}(S) &=& \frac{2^{18} S^4}{3^6 \pi^3} \mathrm{exp} \left( - \frac{64 S^2}{9 \pi}\right). \label{wigner4} 
\end{eqnarray}
A Poisson distribution is obtained for the nearest neighbor spacing
for a random matrix spectrum when the system would exhibit classically
regular motion or equivalently is integrable \cite{haake2001}.  In quantum
mechanics this corresponds to sufficiently many conserved quantum
numbers.  The energy levels are unaffected by
the presence of its neighbors.

\begin{figure}
\includegraphics[width=\linewidth]{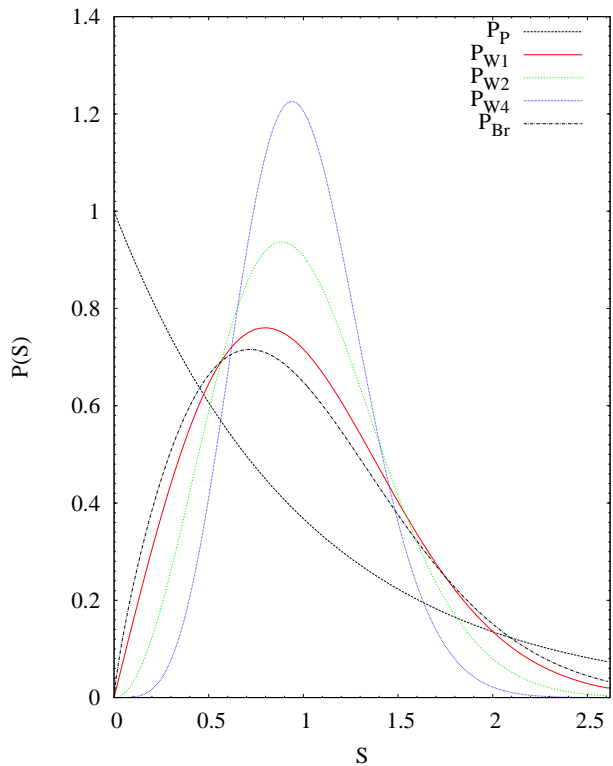}
\caption{The five standard distributions in
  Eqs.\eqref{poisson}-\eqref{brody}. Here $P_P$ is peaked at small
  $S$, and the peaks for the three Wigner distributions move to larger
  $S$-values as the power of $S$ increases. The Brody distribution is 
plotted for a value of the Brody parameter of $\kappa = 0.4$}
\label{fig:distrib}
\end{figure}

On the other hand, the three other distributions obtained from a
random matrix spectrum corresponds to irregular motion where the
levels do not cross.  In this sense the levels repel each other and each
energy level must keep some distance to its neighbors.  This traditionally is the signature of quantum
chaotic motion~\cite{haake2001}.  The three distributions are similar
but with different powers $\beta$ on $S$ corresponding to different
degrees of level repulsion. Indeed, when the neighboring energy levels
come close together (i.e. $S \to 0$) the distributions are defined by
$P(S) \sim S^{\beta},$ where $\beta = 1, 2, 4$.  Clearly an increasing
$\beta$ increases the size of the small $S$ region of vanishing
probability as seen in Fig. \ref{fig:distrib}.

  All the present statistical analyses are based on random
  matrix theory \cite{reichl1992, mehta2004}, where matrix properties
  are analyzed under various assumptions of inherent symmetries.
  Antiunitarity, like time-reversal symmetry, restricts the form of
  the hamiltonian and the corresponding matrix properties, and defines
  the power, $\beta$, of the level repulsion $\beta$ \cite{haake2001}.
  There are three universal degrees of level repulsion: The $\beta =
  1$ case corresponds to the systems with integer spin and invariant
  under time-reversal and with some spatial rotational and/or
  reflection symmetry.  The half-integer spin particles with time reversal
  symmetry and broken geometric symmetries correspond to $\beta = 4$.
  We have an interest in the latter symmetry because it corresponds to
  deformed systems with the Rashba spin-orbit interaction
  \cite{kravtsov2012, haake2001}.  If the time-reversal symmetry is broken by
  e.g. a  magnetic field, then the system is described by $\beta = 2$.

The transition between the Poisson, Eq.\eqref{poisson}, and linear Wigner,
Eq.\eqref{wigner1}, distributions can be described by the so-called
Brody distribution \cite{brody1981, santos2010}, which is expressed as
\begin{equation}
\label{brody}
P_{Br}(S) = (\kappa + 1) b S^{\kappa} exp \left ( -b S^{\kappa + 1} \right ),
\end{equation}
where $\kappa$ ($0 \leq \kappa \leq 1$) is the Brody parameter and $b
= \left [ \Gamma \left ( \frac{\kappa + 2}{\kappa + 1} \right )
  \right]^{\kappa + 1}$. It is shown in Fig. \ref{fig:distrib} for an
intermediate $\kappa$-parameter, and it is straightforward to see the two
limits of Poisson \eqref{poisson} and Wigner \eqref{wigner1}
distributions obtained for $\kappa = 0$ and $\kappa = 1$,
respectively.  The physical meaning of the Brody parameter $\kappa$
can be defined only in particular cases~\cite{sakhr2005}.

\subsection{System specifications}
We investigate a system of spin-orbit coupled non-interacting
identical particles of mass $m$ trapped in a
harmonic potential. The spin-orbit coupling term depends only on
$x$- and $y$-coordinates, and the motion in the $z$-direction can be easily decoupled.
Hence, we will only consider the 2D hamiltonian.
We also consider the Zeeman term
which, for example, can appear by applying an external magnetic field.  
The one-body
hamiltonian for the system is given by
\begin{eqnarray}
\label{eq:hamiltonian}
\hat H &=& \left ( \frac{\mathbf {p^2}}{2m} + \frac{1}{2} m 
(\omega_x^2 x^2 + \omega_y^2 y^2) \right ) \otimes \hat I \nonumber \\ 
&+& \alpha_R ({\hat \sigma_{x}} {p_{y}} - {\hat \sigma_{y}} {p_{x}}) 
+ h_z \hat \sigma_z,
\end{eqnarray}
where $\mathbf{p} = \{p_{x}, p_{y}\}$ and $\mathbf{r} = \{x, y\}$ are
the 2D momentum and coordinate operators, $\omega_x$ and $\omega_y$
are the frequencies of the harmonic oscillator trap in the directions
$x$ and $y$, $\hat I$ is the $2 \times 2$ unit matrix. 
The spin-orbit coupling term is given in a Rashba-like form~\cite{rashba1984}, 
$\alpha_R (p_x \hat \sigma_y - p_y \hat \sigma_x)$, where $\alpha_R$ is
the strength of the spin-orbit coupling, $\hat \sigma_x$, $\hat
\sigma_y$ and $\hat \sigma_z$ are the $2 \times 2$ Pauli matrices and
$h_z\hat \sigma_z$ is the Zeeman term. 

The hamiltonian~\eqref{eq:hamiltonian} without the Zeeman term commutes
with the time reversal operator, which for spin-$\frac{1}{2}$
particles can be written as
\begin{equation}
\hat T = i \hat \sigma_y K,
\end{equation}
where $K$ is the complex conjugation operator~\cite{sakurai1994}.  For
this system the eigenenergies are two-fold (Kramers) degenerate.  For
the case of equal frequencies ($\omega_x =\omega_y$) the system is
cylindrically symmetric.  The deformation of the confining harmonic
potential ($\omega_x \neq \omega_y$) breaks the cylindrical rotational
invariance.  In this case we have a reason to believe that the level
repulsion should be rather strong and the Wigner distribution with
$\beta = 4$ could arise.  If we add an additional external Zeeman
field the time-reversal symmetry is also broken and the distribution
with $\beta = 2$ is suggested.

We emphasize that it is not at all necessary to arrive at one of the
distributions in Eqs.~\eqref{poisson}-\eqref{brody}. Any combination
or totally different results are perhaps more likely. However, given
sets of the tunable parameters may produce these standard
distributions, and thereby provide hamiltonians corresponding to pure
regular or chaotic motion.  We shall therefore perform the appropriate
statistical analysis of the energy spectra in order to understand
these features better.

 The eigenstates of the hamiltonian~\eqref{eq:hamiltonian}
  were previously calculated using the exact diagonalisation
  method~\cite{marchukov2013}, which we adopt for the present paper.
  We expand the eigenstates on a basis of harmonic oscillator wave
  functions.  The oscillator potential is deformed and clearly
  suggesting a corresponding set of basis functions.  However, the
  spin-orbit coupling is then far from diagonal and relatively
  high-lying states can be expected necessary.  We still choose the
  deformed Cartesian oscillator basis of the oscillator trap.  The
  energy units, or equivalently the spatial extension, may be very
  different for the $x$ and $y$-directions.  To account roughly for
  the deformation effect in the basis, we include all states up to the
  same maximum energy, $E_{max} \approx N_x\omega_x \approx N_y
  \omega_y$, where $N_x$ and $N_y$ are the largest directional
  Cartesian quantum numbers.

In the statistical analyses we have to choose the energy window of
eigenvalues such that the number of levels are neither too small nor
too large. Too few levels do not allow proper statistics and too many
emphasize the basis properties instead of the interactions in the
hamiltonian.  To be clear about this, a large number of levels may all
be found numerically as precise eigenvalues, but inclusion of all such
high-energy solutions in an analysis would not provide useful
dynamical information about the behavior of the system.  The most
appropriate energy interval depends on the interactions. Our choice of
moderate spin-orbit strengths allow a reasonable guess of energy
window independent of choice of the controllable parameters.  In this
paper we analyse usually the lowest $150$ doubly degenerate levels,
while they are obtained by any basis with more than $700$
correspondingly degenerate states.  The average maximum quantum number
in each direction is then for $700$ states about $26$. Then all states
in the analyses are fully converged.

The calculated energy spectra depend on the three parameters: the
spin-orbit coupling strength, $\alpha_R$, the deformation,
$\gamma=\omega_x/\omega_y$, of the trap, and the Zeeman strength,
$h_z$.  Since essentially all symmetries are broken for general values
of the deformation, we can expect the repulsive behavior between the
energy levels seen as avoided crossings. This might be a signature of
chaos in the system, and the Zeeman term might possibly be used to
move between the quadratic and quartic Wigner distributions.

\subsection{Statistical treatment}
The set of eigenvalues, $\{\varepsilon_i\}$, derived from the
Schr\"{o}dinger equation with the hamiltonian~{\eqref{eq:hamiltonian}}
now has to be analysed.  First, we note that for the analysis of nearest neighbor
distributions systematic degeneracies can be removed, since they only
add points of zero spacing. For the present case, the system with $h_z=0$
obeys time-reversal symmetry, and in that case
all levels are doubly degenerate where only one set has to be
treated.
Second, the analysis in terms of dimensionless distributions  
requires removal of the scale carrying the unit of energy. In
addition, it is also necessary to even out the average scales
of the spacings between the nearest energy levels arising from the  
different density of the energy spectrum.

We introduce the staircase function:
\begin{equation}
\sigma(\varepsilon) = \frac{1}{N} \int_{-\infty}^\varepsilon\sum_{i=1}^N \delta(\varepsilon' - \varepsilon_i) \mathrm{d}\varepsilon',
\end{equation}
where $\varepsilon$ is an energy value, $N$ is the number of levels
included in the analysis, and $\delta$ is the Dirac delta
function. This function calculates the number of available levels below a given energy and
carries the properties of the single-particle level spectrum used in our statistical analyses.

To smooth out $\sigma(\varepsilon)$ we substitute each delta function
by a continuous normalized distribution centered at the same energy.
For calculational simplicity we choose normalized Gaussians, hence obtaining
\begin{eqnarray}
 \overset{-}{\sigma}_{\Delta}(\varepsilon) = \frac{1}{\Delta \sqrt{\pi}} \frac{1}{N} \int_{-\infty}^\varepsilon \mathrm{d}\varepsilon'
\sum_{i=1}^N e^{- \left (\frac{\varepsilon' - \varepsilon_i}{\Delta} \right )^2} \nonumber \\
\times \left (\frac{15}{8} - \frac{5}{2} \left ( \frac{\varepsilon' - \varepsilon_i}{\Delta} \right )^2 + \frac{1}{2} \left (\frac{\varepsilon' - \varepsilon_i}{\Delta} \right ) ^4 \right ) , 
\end{eqnarray}
where the smearing parameter, $\Delta$, appears as the width of the
Gaussians. The fourth order polynomial guarantees that any initial
smooth behavior reproducible by such a polynomial correctly reappears
after the smoothing~\cite{brack1972}. Different polynomial orders can
be chosen but fourth order is sufficient to provide a reasonable
stability range in $\Delta$, where
$\overset{-}{\sigma}_{\Delta}(\varepsilon)$ in practice is essentially
independent of $\Delta$.

Finally, to obtain a spectrum with an average level spacing normalized
to one, we map the spectrum $\{\varepsilon_i\}$ onto a new spectrum,
$\{e_i\}$, defined by
\begin{equation}
e_i = N \overset{-}{\sigma}_\Delta(\varepsilon_i),
\end{equation}
for $i = 1, \dotsc, N$. This procedure is usually called {\it
unfolding of the spectrum}~\cite{haake2001}. The statistical
analysis are now applied on the new scale-independent and
dimensionless spectrum, $\{e_i\}$.

It is not surprising that for reasonable statistical analysis
the number of levels $N$ should be "big enough". The size of it
is of course always specific to a problem at hand. The rule of thumb
is to make sure that the levels are more or less evenly distributed even
before the unfolding of the spectrum. In that case the unfolding works the
best and the statistical properties of the spectrum are the most evident.

\section{Numerical results}\label{sec:results}
To obtain the energy spectra we use the exact diagonalization method
described in Ref.~\cite{marchukov2013}.  We vary both the strengths,
$h_z$ and $\alpha_R$, of the Zeeman and spin-orbit coupling terms, and
the deformation of the trap expressed as the ratio of frequencies,
$\gamma = \omega_x/\omega_y$, in $x$ and $y$ directions.  We
  focus mostly on the spin-orbit strength which therefore is used as
  the parameter in all figures. We confine ourselves to strengths
  varying from zero to the natural value in oscillator units,
  $\sqrt{\hbar\omega_y/(2m)}$.  For much larger values the spin-orbit would
  be dominating and the Landau-like levels would appear.  We want to
  investigate the interplay between the deformed oscillator trap and
  the spin-orbit coupling. Therefore we choose these terms of comparable
  size.  

The number of levels per unit energy increases with energy and the
lowest are usually less significant due to their relative small
numbers. However, they may also exhibit a completely different
structure as exemplified by a large deformation where 1D structure
appears at the bottom of the spectrum and 2D is restored at energies
higher than $\hbar$ times the perpendicular frequency.  This
  dependence on energy window is important.  We select one
  intermediate energy cutoff which is characteristic for the results.
  At the end of this section we discuss briefly the level distributions
  for different windows.

\begin{figure}
\includegraphics[width=\linewidth]{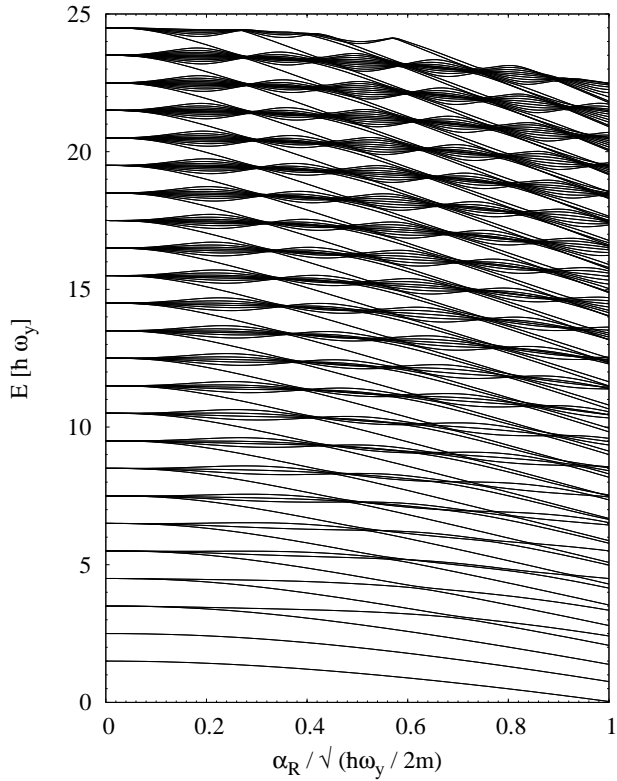}
\caption{The energy levels of a spin-orbit coupled particle with $h_z=0$ in a deformed harmonic trap as a function of spin-orbit coupling strength. The deformation is $\gamma = 2$. The levels from 0 to 300 are shown (notice that levels are two-fold degenerate).}
\label{fig:spectrum1}
\end{figure}

\subsection{Implications of symmetries}
We first must identify possible symmetries, remove systematic
degeneracies, separate eigenvalues and eigenstates in sets corresponding to conserved quantum
numbers, and treat each decoupled set independently. The 2D oscillator
alone is known to have a series of degeneracies for integer values of $\gamma$, 
and the smallest integers have largest degeneracy. These real crossings 
(not avoided) are present because the uncoupled levels are defined 
by the conserved oscillator quanta in the two directions. The degeneracies 
are connected to the classical orbits defined as closed paths by successive
reflection on the potential walls \cite{brack1997}. 

Close to these deformations with special oscillator degeneracy there
must be many close-lying levels due to the small degeneracy breaking.
In addition the other nearest neighbors, above and below each of these
near-degenerate levels, are much further apart.  Thus, we can expect
an unfolded distribution with peaks at small distance and a peaked
distribution at distances larger than the average from 
the spectral theory of random matrices.

\begin{figure}
\includegraphics[width=\linewidth]{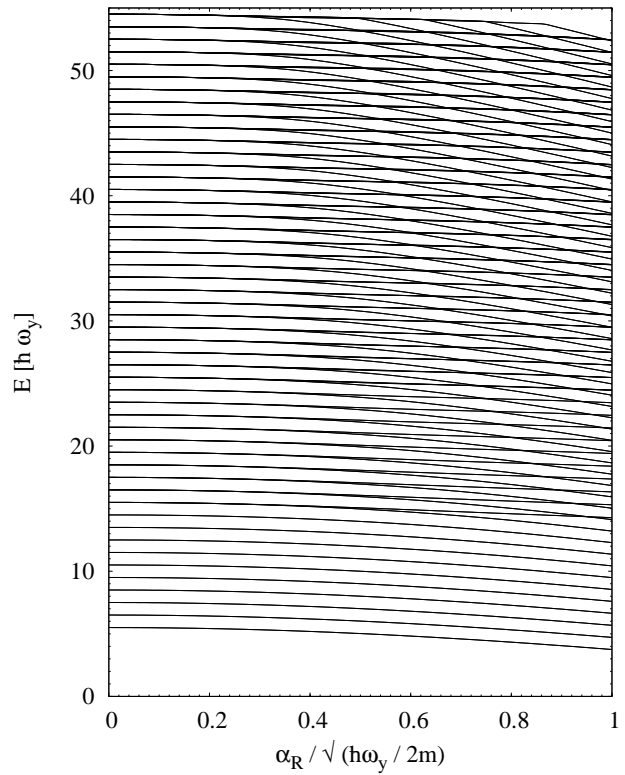}
\caption{The energy levels of a spin-orbit coupled particle with $h_z=0$ in a deformed harmonic trap as a function of spin-orbit coupling strength. The deformation is $\gamma = 10$. The levels from 0 to 300 are shown (notice that levels are two-fold degenerate)}
\label{fig:spectrum2}
\end{figure}

Addition of the spin-orbit coupling lifts the oscillator degeneracy
and destroys the related symmetries. Level couplings are now present,
conserved quantum numbers essentially disappeared, and the remaining
time-reversal symmetry for $h_z=0$ only produces doubly degenerate levels. The energy levels are
now prohibited to cross and the spectrum appears to have a "band structure".

The origin of these "bands" is intuitively quite clear: the spin-orbit
coupling lifts the oscillator degeneracy in a way not unlike, for
instance the Zeeman splitting of levels. The avoiding crossing, on the
other hand, also controls the behavior of levels and leads to the
opening and closing of the "gaps" and, thus, to the "band structure".

\begin{figure}
\includegraphics[width=\linewidth]{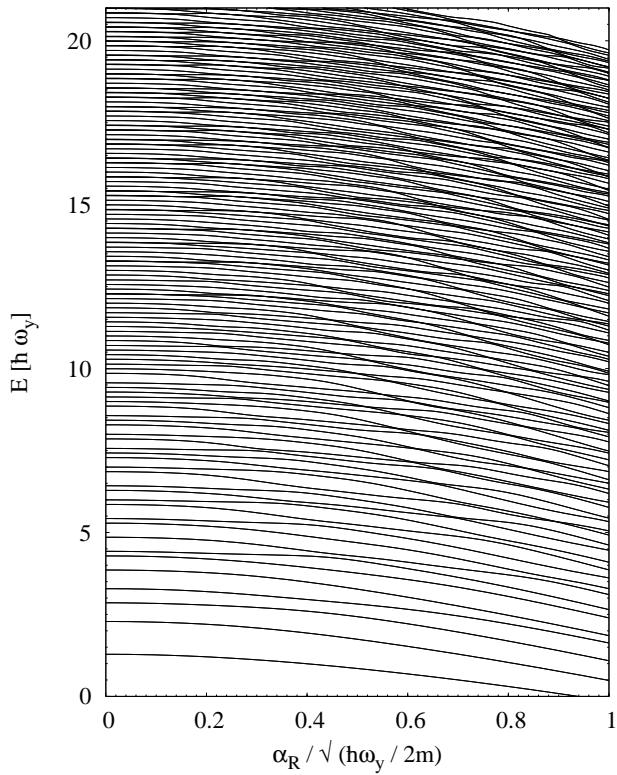}
\caption{The energy levels of a spin-orbit coupled particle with $h_z=0$ in a deformed harmonic trap as a function of spin-orbit coupling strength. The deformation is $\gamma = 1.57$ (left). The levels from 0 to 300 are shown (notice that levels are two-fold degenerate).}
\label{fig:spectrum3}
\end{figure}

The deformation affects this structure. The deformed oscillator levels
are shifted compared to non-deformed ones and the degeneracies might
be lifted.  However, for integer frequency ratios, energy levels
(except for the lowest ones) are still degenerate, because in this
case they are shifted by an integer number of oscillator quanta. This
means that the "band structure" does not disappear, but still exists
in the spectrum in the presence of the spin-orbit coupling. This is
illustrated in Fig.~\ref{fig:spectrum1} for the frequency ratio of
$\gamma = 2$.

An extreme case of large frequency ratio of $\gamma = 10$ is shown in
Fig.~\ref{fig:spectrum2}. The same pattern of degeneracy restoring as
for $\gamma = 2$ is seen but now appearing only for much larger
strengths, $\alpha_R$, and at higher energies. The 1D limit is efficiently
established for the lowest levels corresponding to energies where the
 excitations in $x$-direction are almost forbidden. We then conclude that
destruction of symmetries or lifting degeneracies by using non-integer
frequency ratios seems to be as effective both with and without
spin-orbit coupling, although additional interaction between 
levels must tend to randomize better.

\begin{figure}
\includegraphics[width=\linewidth]{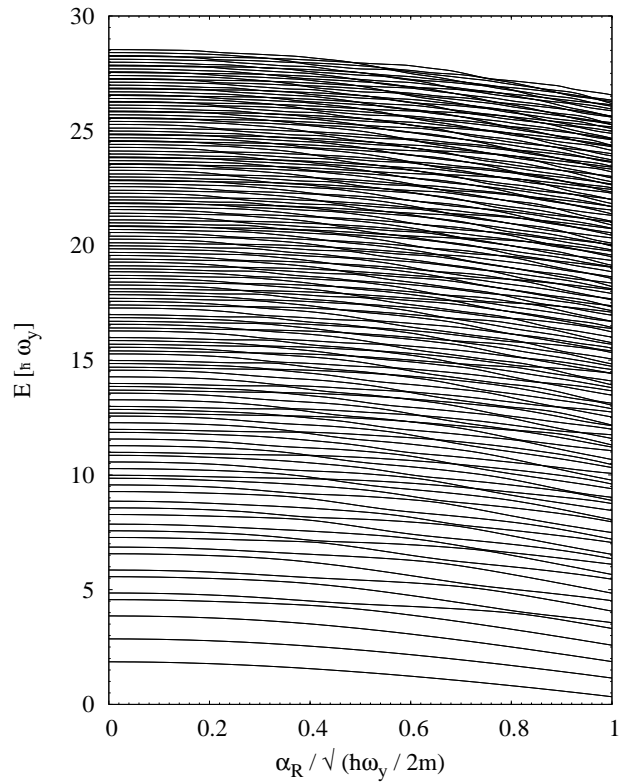}
\caption{The energy levels of a spin-orbit coupled particle with $h_z=0$ in a deformed harmonic trap as a function of spin-orbit coupling strength. The deformation is $\gamma = 2.71$. The levels from 0 to 300 are shown (notice that levels are two-fold degenerate).}
\label{fig:spectrum4}
\end{figure}

The energy levels should be much more evenly distributed
for non-integer frequency ratios where oscillator degeneracies are much more rare. 
In Figs.~\ref{fig:spectrum3} and \ref{fig:spectrum4} we show spectra for $\gamma =1.57$ and
$\gamma = 2.71$ where the latter is rather far from any ratio of very
small integers, the closest are $8/3$ and $11/4$. The white regions
("gaps") are still clearly visible.  However, the structures are now
much weaker and apparently almost vanishing for the highest energies
in the spectrum.

\begin{figure}
\includegraphics[width=\linewidth]{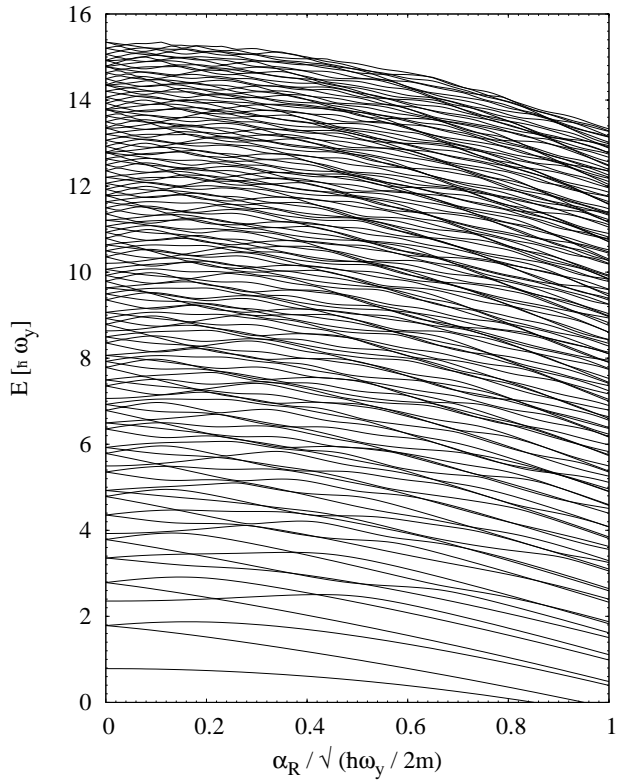}
\caption{The energy levels of a spin-orbit coupled particle in a
  deformed harmonic trap as a function of spin-orbit coupling
  strength. The deformation $\gamma = 1.57$. The Zeeman strength is $h_z=0.5\hbar\omega_y$.
  The levels from 0 to 150 are shown.}
\label{fig:spectrum5}
\end{figure}

\begin{figure}
\includegraphics[width=\linewidth]{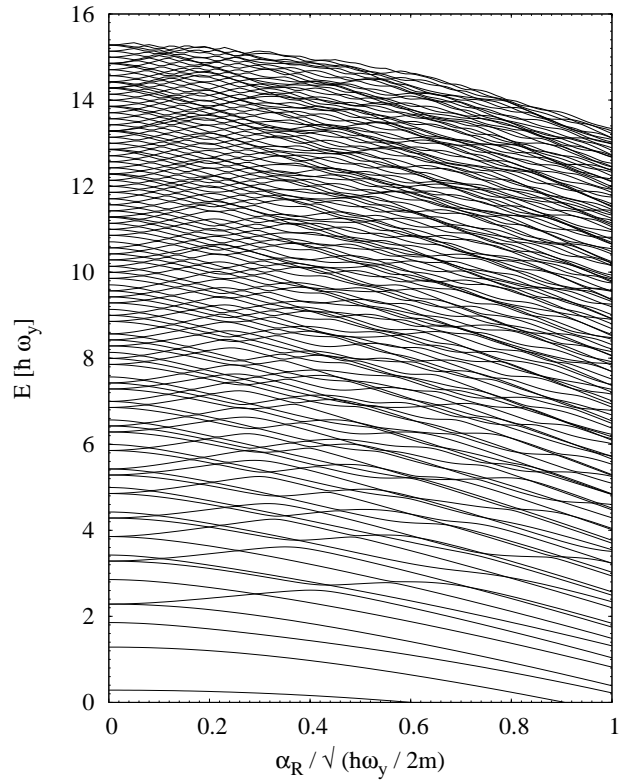}
\caption{The energy levels of a spin-orbit coupled particle in a
  deformed harmonic trap as a function of spin-orbit coupling
  strength. The deformation $\gamma = 1.57$. The Zeeman strength is $h_z=\hbar\omega_y$.
  The levels from 0 to 150 are shown.}
\label{fig:spectrum6}
\end{figure}

So far we omitted the Zeeman field which may introduce new features
since it breaks the Kramers degeneracy.
In Figs.~\ref{fig:spectrum5} and \ref{fig:spectrum6} we show spectra
with finite Zeeman field and the broken rotational symmetry. The
number of levels now doubled due to breaking the time-reversal
symmetry. The denser spectra have now only avoided level crossings
even though several levels appear to be very close-lying.  In any case
the degree of level repulsion cannot be read directly off from these plots.

The repulsion between energy levels is always present in some limited
regions of energy for corresponding deformations and interaction
strengths. From the presented level spectra it is clear that a given
hamiltonian with both finite deformation and spin-orbit interaction
only lead to occasional avoided crossing regions in rather small
energy windows.

 A special type of symmetry is releated to the spin-orbit
  coupling via the balance of the two terms, $\hat \sigma_y p_x$ and
  $\hat \sigma_x p_y$.  Varying their relative strengths continuously
  from $-1$ to $1$ change the spectra substantially. The symmetry
  between the $x$ and $y$ directions implies that the Rashba ($-1$)
  and Dresselhaus ($1$) spin-orbit couplings produce identical
  spectra.  When one of the spin-terms is removed the hamiltonian
  decouples into two $1D$ contributions, where each roughly speaking
  is a shifted oscillator. Thus, such parameter variations change from
  Rashba coupling to an effective $1D$ hamiltonian.  The intermediate situation when the spin-orbit
  coupling strengths are not equal is usually referred as the anisotropic
  spin-orbit coupling in a spherical $2D$ trap. The hamiltonian is then equivalent
  to that of the Jahn-Teller model \cite{larson2009}. This model and it relation
  to the irregular dynamics have been studied in Refs.~\cite{markiewicz2001,yamasaki2003, majernikova2006a,
  majernikova2006b}.
  While we will not consider anisotropic spin-orbit terms in this
  paper we compare and contrast the present study to previous 
  work on the Jahn-Teller model.

\begin{figure}
\centering	
\includegraphics[width=\linewidth, height=4.3cm]{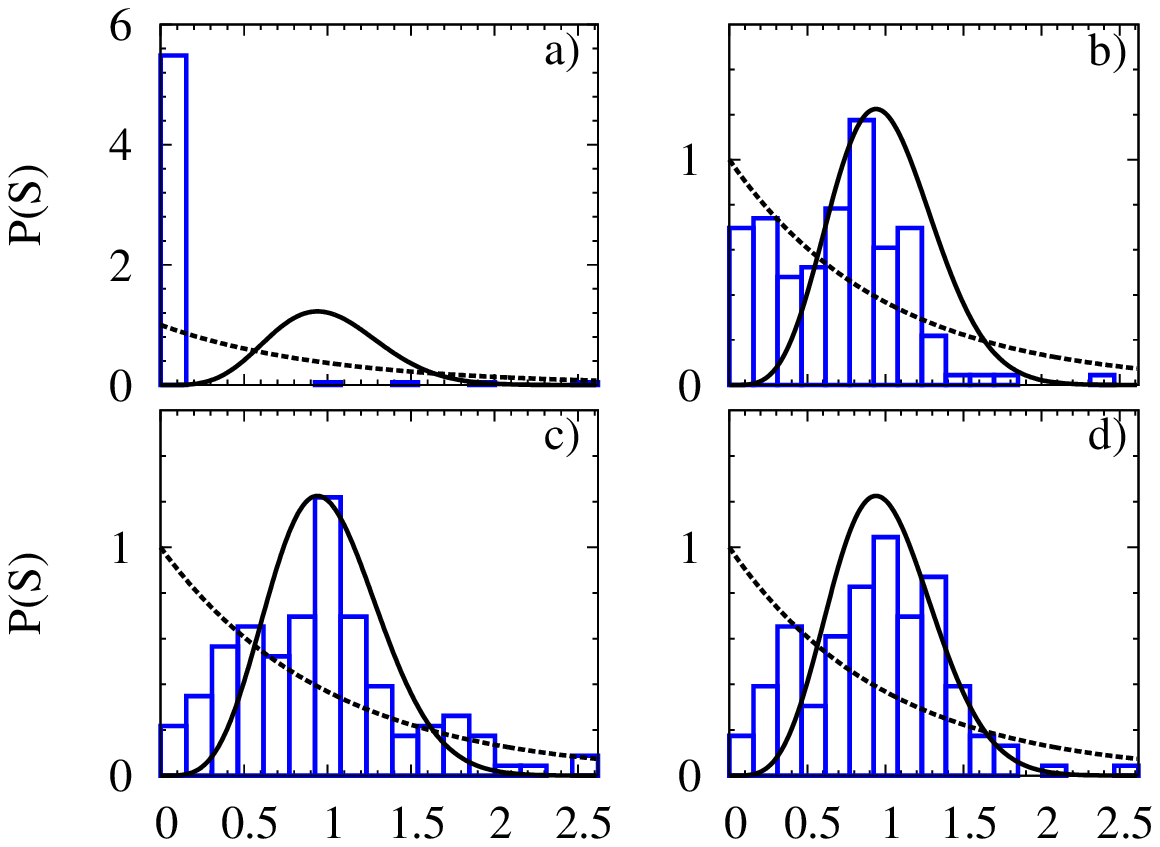}
\includegraphics[width=\linewidth, height=4.3cm]{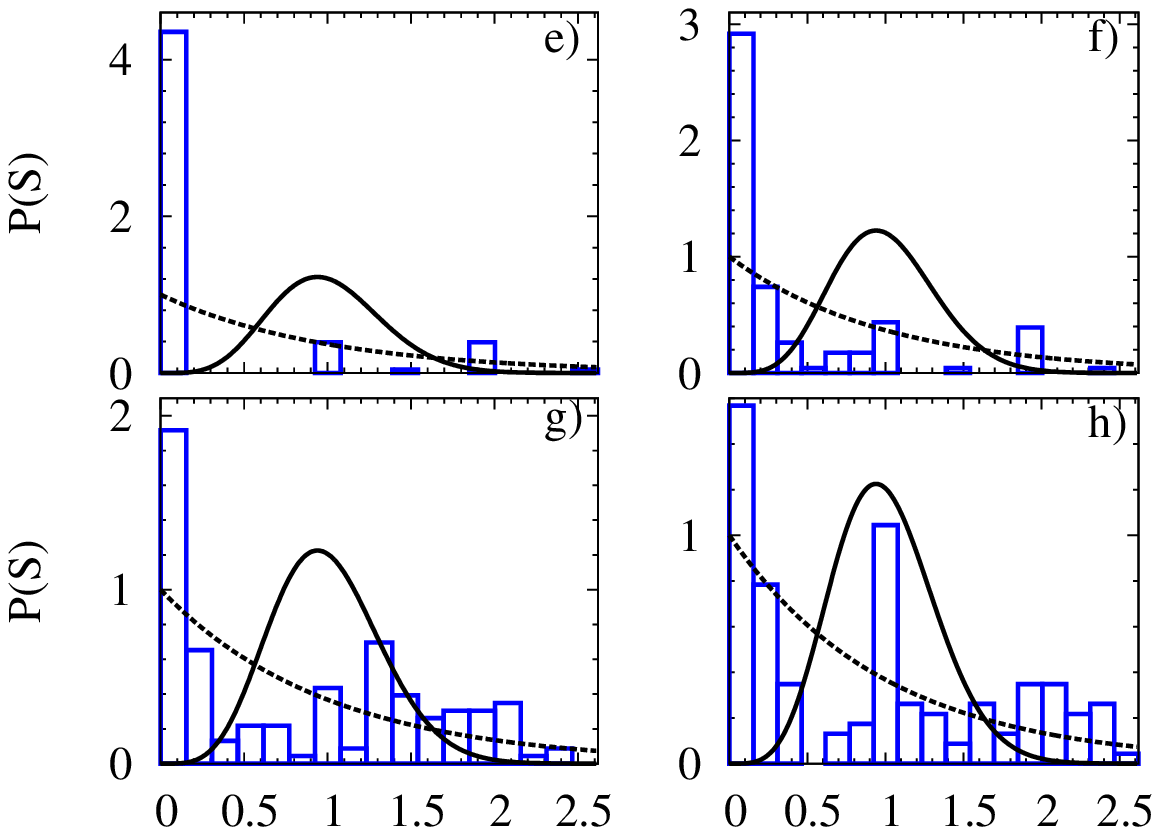}
\includegraphics[width=\linewidth, height=4.3cm]{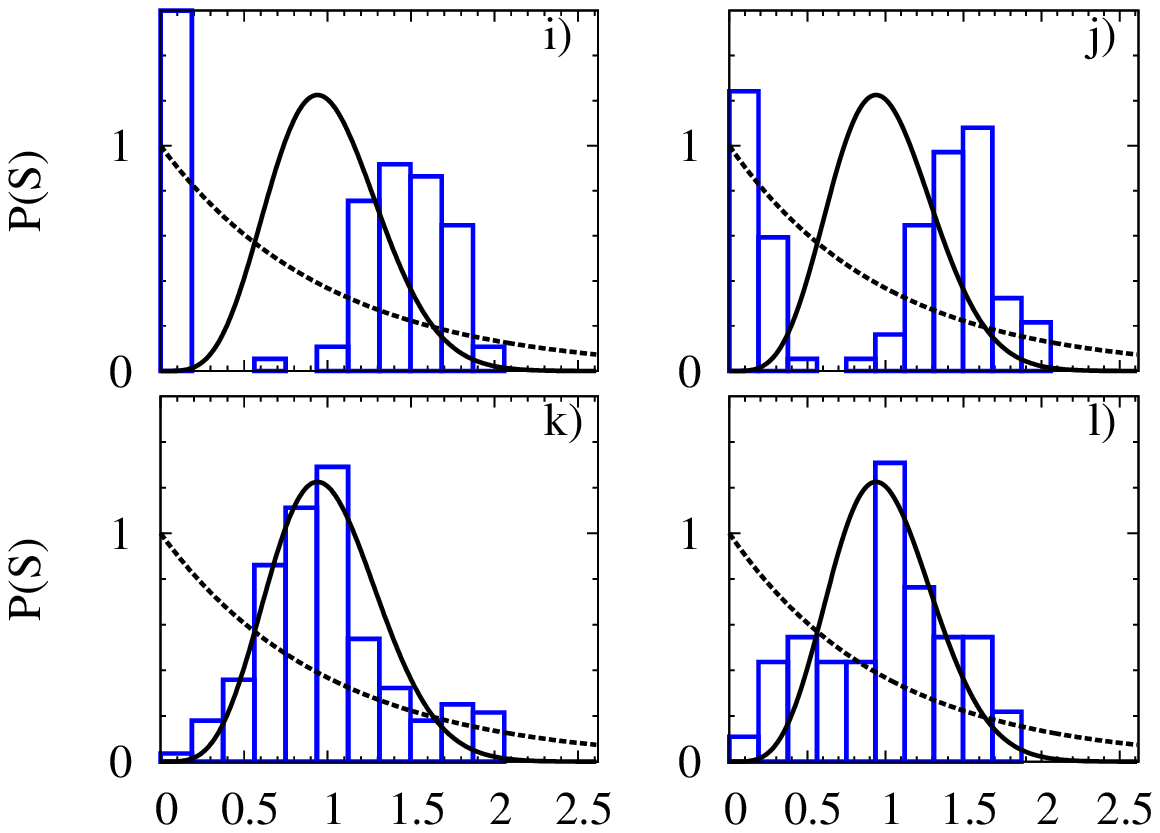}
\includegraphics[width=\linewidth, height=4.3cm]{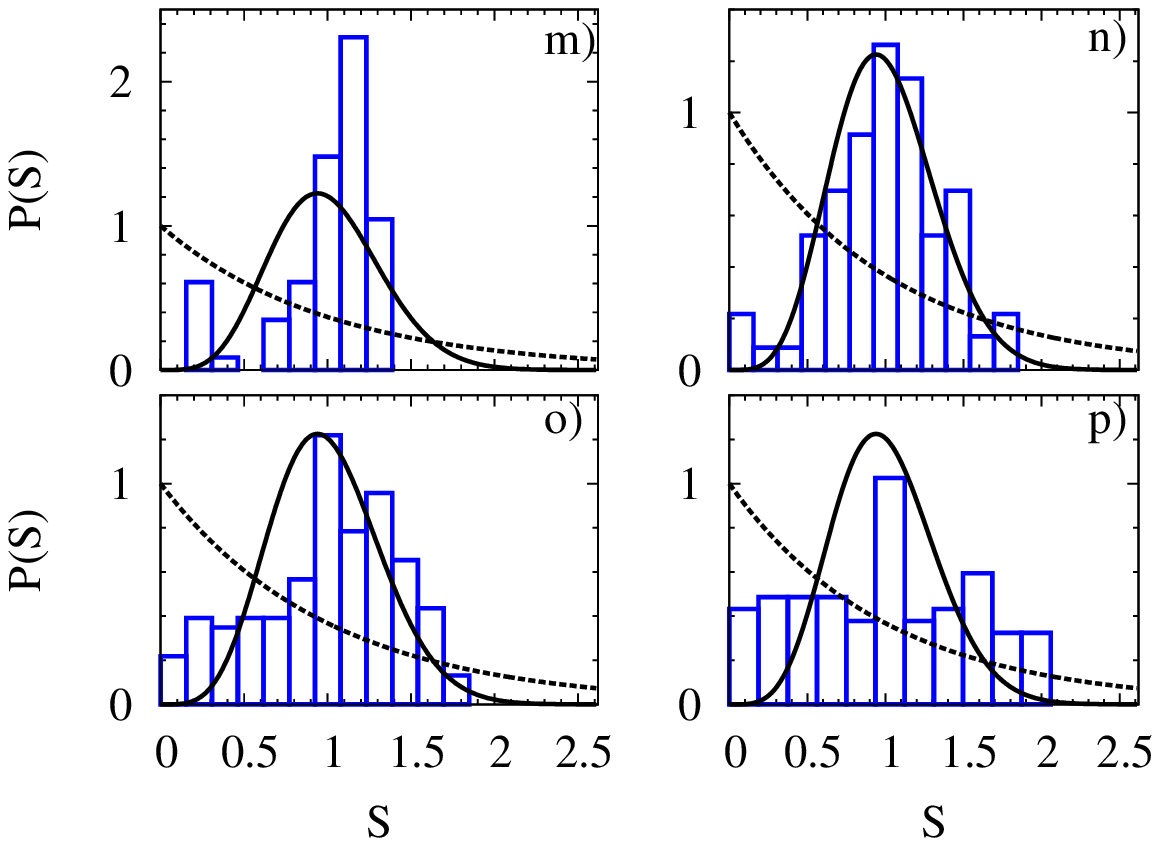}
\caption{The nearest neighbor level spacing distribution for different sets of parameters values. For deformation $\mathbf{\frac{\omega_x}{\omega_y} = 2}$: a) $\alpha_R = 0$, b) $\alpha_R = 0.3$, c) $\alpha_R = 0.5$, and d) $\alpha_R = 0.7$. For deformation $\mathbf{\frac{\omega_x}{\omega_y} = 10}$: e) $\alpha_R = 0$, f) $\alpha_R = 0.3$, g) $\alpha_R = 0.5$, and h) $\alpha_R = 0.7$. For deformation $\mathbf{\frac{\omega_x}{\omega_y} = 1.57}$: i) $\alpha_R = 0$, j) $\alpha_R = 0.3$, k) $\alpha_R = 0.5$, and l) $\alpha_R = 0.7$. For deformation $\mathbf{\frac{\omega_x}{\omega_y} = 2.71}$: m) $\alpha_R = 0$, n) $\alpha_R = 0.22$, o) $\alpha_R = 0.3$, and p) $\alpha_R = 0.5$. We use units $\sqrt{\frac{\hbar \omega_y}{2m}}$ for the spin-orbit coupling parameter $\alpha_R$. The external magnetic field $h_z=0$. The distributions are compared to the Poisson (dashed line) and Wigner \eqref{wigner4} (solid line) distributions.}
\label{fig:histograms1}
\end{figure}

\begin{figure}
\centering	
\includegraphics[width=\linewidth]{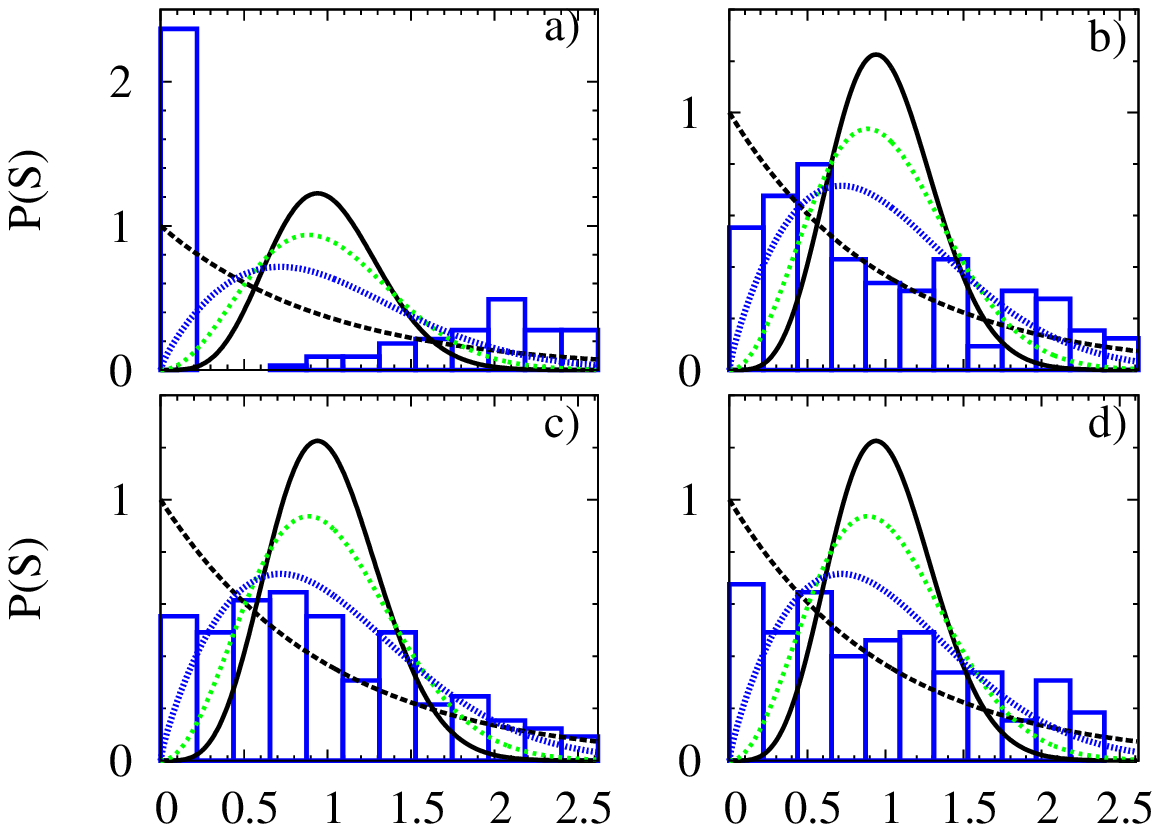}
\includegraphics[width=\linewidth]{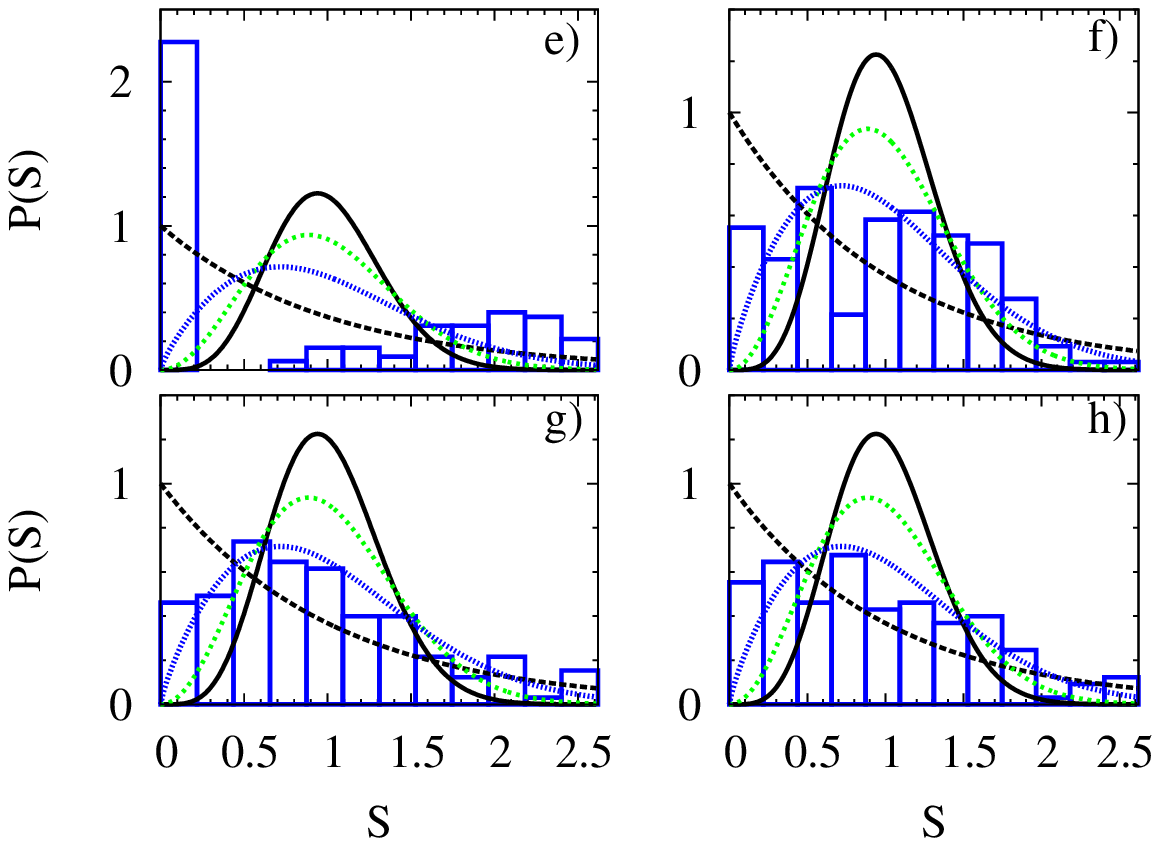}
\caption{The nearest neighbor level spacing distribution for different values of spin-orbit coupling parameter, $\alpha_R$, in units of $\sqrt{\frac{\hbar \omega_y}{2m}}$: a) and e) $\alpha_R = 0$, b) and f) $\alpha_R = 0.3$, c) and g) $\alpha_R = 0.5$, d) and h) $\alpha_R = 0.7$. The cases from a) to d) correspond to the Zeeman strength $h_z=\hbar \omega_y/2$ and the cases e) to h) correspond to $h_ z= \hbar \omega_y$. The harmonic trap frequencies ratio $\frac{\omega_x}{\omega_y} = 1.57$. The distributions are compared to the Poisson (dashed line), Brody (solid blue line) \eqref{brody}, quadratic Wigner (solid green line) \eqref{wigner2} distributions, and quartic Wigner (solid black line) \eqref{wigner4} distributions. The Brody parameter is $\kappa = 0.4$.}
\label{fig:histograms2}
\end{figure}

\subsection{Nearest neighbor distributions}
The distributions are the first step to pin-point an underlying
regular or irregular motion.  To determine whether a system is quantum
chaotic or not typically requires an analysis of the
equivalent classical problem~\cite{haake2001,reichl1992}.  Such a full
analysis for the system in an isotropic ($\omega_x = \omega_y$) trap
with anisotropic spin-orbit coupling ($\hat V_{SOC} = \alpha_1 \hat
\sigma_x p_y + \alpha_2 \hat \sigma_y p_x,$ with $\alpha_1 \neq
\alpha_2$) was recently published in Ref.~\cite{larson2013}. The analysis
presented in that paper can be carried out for our system as well and we
assume that the correspondence between classical and quantum chaotic
behaviour holds in our case as well. A very interesting recent study
of quantum billiard with spin-orbit coupling \cite{khomitsky2013} 
provides further evidence that chaos will arise for spin-orbit coupled
systems in two dimensions.

Within the context of the Jahn-Teller model, the
level distributions statistics have been considered in 
previous works \cite{yamasaki2003,
majernikova2006a, majernikova2006b}. The Hamiltonian 
in those papers can be mapped to isotropic and 
anisotropic spin-orbit couplings in a spherical 
$2D$ harmonic oscillator in our setup. Furthermore, 
a Jahn-Teller equivalent of a deformed oscillator 
discussed in Ref.~\cite{markiewicz2001} was shown 
to have signatures of chaos in the quantum dynamics
although no level distributions were considered. 
In Ref.~\cite{yamasaki2003} it was suggested that 
anharmonic (oscillator) terms could be used to 
drive the level distribution from Poisson \eqref{poisson}
to Wigner \eqref{wigner1} and that such anharmonicity 
could be a requirement for chaotic behaviour. 
However, subsequent studies in Refs.~\cite{majernikova2006a,majernikova2006b}
have found substantial deviations from the Poisson \eqref{poisson} 
distribution without including anharmonicity. 
In Ref.~\cite{majernikova2006a} these deviations are
analyzed using distributions different from Wigner \eqref{wigner1}.
The results presented below also go beyond the Poisson \eqref{poisson}
and Wigner \eqref{wigner1} distributions. By using a combination 
of Rashba spin-orbit coupling, and deformed trap and a Zeeman field
we find parameter regimes where the Wigner \eqref{wigner4} and
the Brody \eqref{brody} distributions may be realized. 
While the Brody distribution has been discussed in the 
context of the Jahn-Teller model previously \cite{majernikova2006a},
the fourth order Wigner \eqref{wigner4} was not 
observed as far as we can tell.

Apart from the extreme limits of huge deformation or huge spin-orbit
strengths, we expect the spectra of a $2D$ deformed
oscillator to have different properties due to the breaking of the
cylindrical symmetry seen in the non-deformed case.  
The oscillator degeneracies are varying very
strongly from spherical to well deformed.  For a small number of
rational frequency ratios the degeneracies are largest, but for
irrational frequency ratios the degeneracies have effectively
completely disappeared. This is directly related to properties of the
level spectra and therefore decisive for underlying regular or chaotic
motion.

We shall first  only examine the statistical behavior as a function of
deformation and spin-orbit strength with $h_z=0$.  The analysis provides nearest
neighbor distributions which we shall present as histograms derived
from the unfolded spectrum. As references we compare the results with
the extremes of the Poisson and Wigner distributions Eqs.~\eqref{poisson}
and \eqref{wigner4}.
We avoid the most symmetric case of the spherical oscillator where the
regular structure and large degeneracy initially prohibits a
meaningful statistical analysis. Still for a frequency ratio of $2$
without spin-orbit, see Fig.~\ref{fig:histograms1}a, the level
distribution is trivial with many nearest neighbors of either zero
distance or reflecting the regularity of the oscillator
degeneracy. These features change as the spin-orbit coupling is
switched on. For all finite strengths shown in
Fig.~\ref{fig:histograms1} we find structures resulting from incoherent
and sizewise comparable contributions of the Poisson, $P_P$, and Wigner,
$P_{W4}$, distributions.

The other special case of deformation is shown in
Fig.~\ref{fig:histograms1}(e-h) for a frequency ratio of $10$. The lowest
part of the spectrum reflects the regularity of a 1D oscillator in
Fig.~\ref{fig:spectrum2}, whereas  the 2D behavior emerges only at higher
energies. For larger strengths and higher energies the spectral pattern resembles a
condensed version of the structure for frequency ratio $2$. The
nearest neighbor distribution still reveals a large peak at small
level distance and irregular contributions for larger level distances.

The lack of symmetries for non-integer frequency ratios leads to the
distributions shown in Figs.~\ref{fig:histograms1}(i-p).  We can see
that in both cases the Wigner  distribution, $P_{W4}$, is realized for
particular values of the spin-orbit coupling strength: 
$\alpha_R / \sqrt{\frac{\hbar \omega_y}{2 m}}= 0.5$ for $\gamma = 1.57$,
Fig.~\ref{fig:histograms1}k, and $\alpha_R / \sqrt{\frac{\hbar
    \omega_y}{2 m}} = 0.21$ for $\gamma = 2.71$,
Fig.~\ref{fig:histograms1}n. It implies that in the vicinity of these
values our system is likely to be in the quantum chaotic regime.

Away from oscillator isotropy the level statistics is quite different.
For small spin-orbit parameter the statistical behavior is defined by
the initial oscillator spectrum.  If the higher oscillator levels are "almost
degenerate" in the sense that the difference between levels much
is much smaller than $\hbar \omega_y$ even after the
deformation, then it will appear on the histogram as a strong peak
near the point $S = 0$, see Fig.~\ref{fig:histograms1}i.  In
Fig.~\ref{fig:histograms1}m one sees a different picture: there is no
peak at $S=0$ due to more evenly distributed oscillator eigenlevels.

Combining spin-orbit coupling and deformation introduce a distribution
altogether different from either spin-orbit or deformation separately. Absence of the quantum
mechanical integrals of motion leads to series of energy levels with
avoided crossings. The spectra can be far from the ordered
oscillator structures.  The regions with few levels are
smaller and much less frequent. The pattern seen in
Figs.~\ref{fig:spectrum1}-\ref{fig:spectrum4} are level structures
where low density regions move depending on energy and spin-orbit
strength. The levels for one strength are therefore much more
uniformly distributed. Finite spin-orbit and deformation do not in
general allow symmetries or degeneracies, although a number of
individual almost crossing levels appear in level diagrams.  However,
they are few and in fact avoided crossings. The NNS distributions are
therefore more likely to follow one of the Wigner distributions.

In Fig.~\ref{fig:histograms2} we show the
NNS distribution of levels in the presence of the Zeeman term.  
We consider the same cases of the trap deformation and the
spin-orbit coupling strength as in Fig.~\ref{fig:histograms1}.  The
Kramers degeneracy is lifted by the Zeeman term and its
strength, $h_z$, is another parameter varying the NNS distribution.  We see
that for non-zero values of the spin-orbit coupling strength the
histograms are reproduced very well by the Brody distribution with a
value of the Brody parameter of $\kappa = 0.4$.  This is somewhat
surprising, since after lifting the Kramers degeneracy
one would intuitively expect better agreement with the
second (not first) power Wigner distribution~\eqref{wigner2} \cite{haake2001}.

\begin{figure}
\centering	
\includegraphics[width=\linewidth]{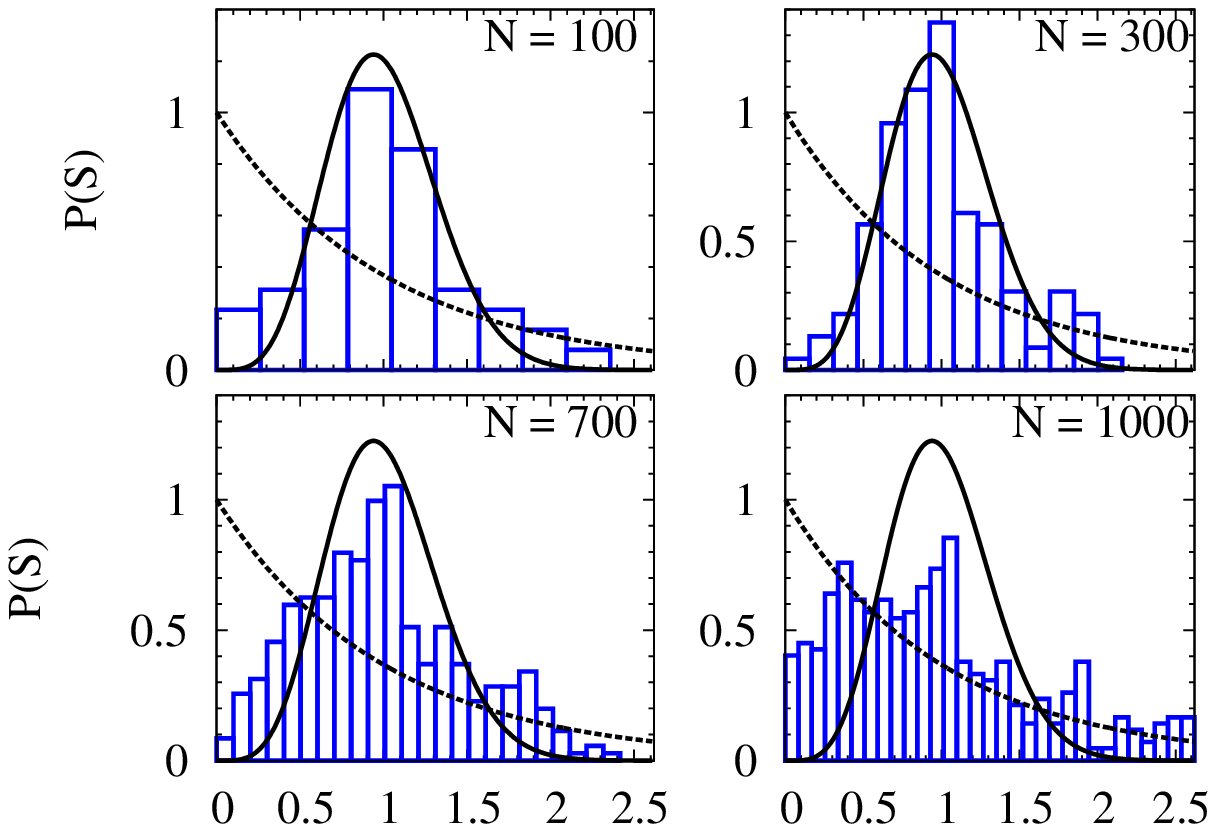}
\includegraphics[width=\linewidth]{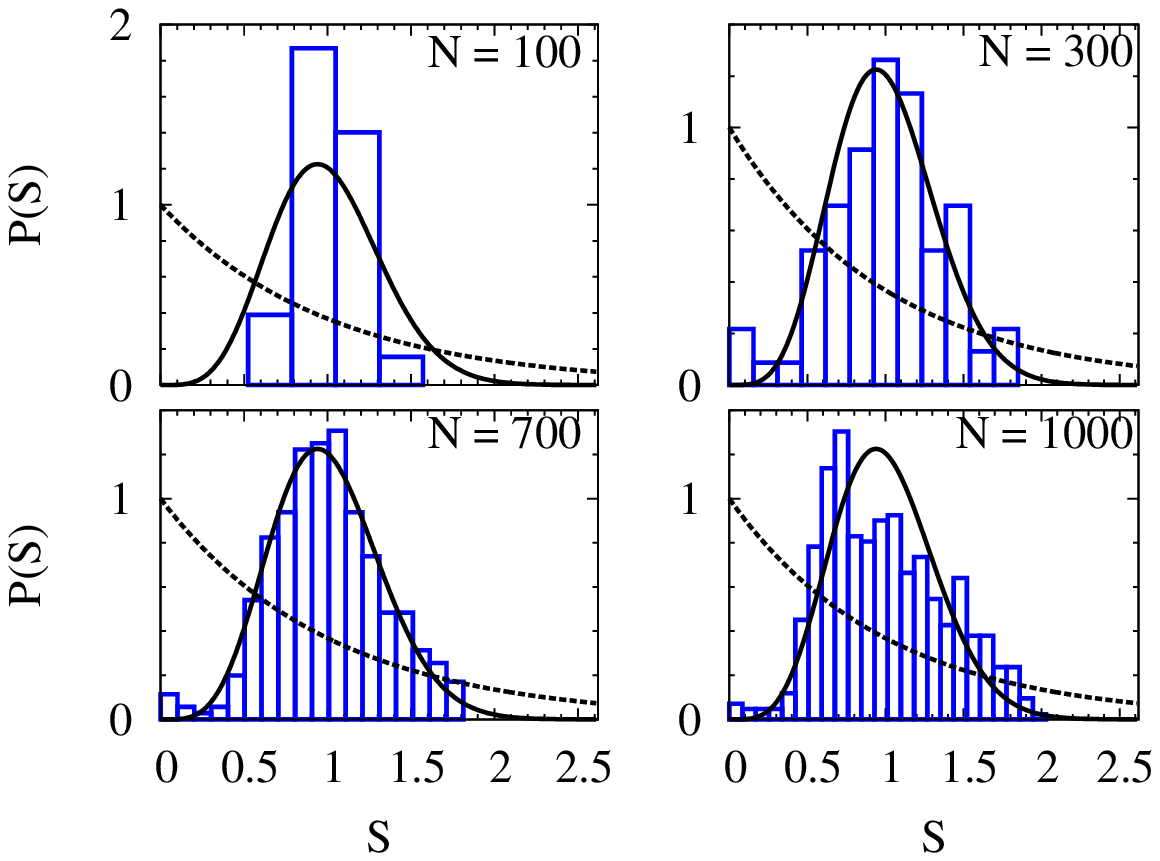}
\caption{The nearest neighbor level spacing distribution for different
  energy windows determined by the number of levels included in the
  analysis. The top four figures are for $\frac{\omega_x}{\omega_y} = 1.57$ and $\alpha_R / \sqrt{\frac{\hbar \omega_y}{2m}} = 0.5$. The bottom four figures are for $\frac{\omega_x}{\omega_y} = 2.71$ and $\alpha_R / \sqrt{\frac{\hbar \omega_y}{2m}} = 0.21$. The number of the lowest energy levels used in the analysis is given in the figures.
The distributions are compared to the Poisson (dashed line) and Wigner \eqref{wigner4} (solid line) distributions.}
\label{fig:histograms3}
\end{figure}

We have so far only shown distributions for $150$ (double or
  single) degenerate levels.  The most interesting variation is to see
  whether the Wigner distribution depends on the energy window. The
  results are shown in Fig.~\ref{fig:histograms3} for two
  distributions corresponding to irregular dynamic motion with $150$
  levels.  The systematic behavior is that an intermediate number of
  levels leave the Wigner distribution essentially unchanged.  Many
  more levels, where the effect of the interaction fade away,
  increases the probability at small distances and the distribution
  would eventually approach a Poisson distribution reflecting very
  regular motion.  Fewer of the low-lying levels must eventually
  describe, perhaps statistically insignificantly, the properties of the
  very small excitations.  By reduction of window size these
  Wigner-like distributions can then move in any direction.

We also changed window size away from $150$ levels for cases where the
Wigner distribution is far from being followed.  The results are
systematically that Wigner-like distributions do not appear when
absent for the intermediate number of levels in the enrgy window.
Thus, we conclude that the $150$ levels used throughout the present
series of analyses is a characteristic number of levels well suited to
search for possible irregular motion.

\section{Conclusions}\label{sec:conclusions}
In this article we discuss the statistical properties of the spectrum
of energy eigenvalues for a two-dimensional one-body hamiltonian with
Rashba spin-orbit and Zeeman terms and a deformed harmonic oscillator.  We
summarized some of the possible
statistical distributions that we may expect for the nearest neighbors
energy separation.  These schematic limiting distributions emerge as
results from random matrix eigenvalues with different symmetries.  In
particular we have regular integrable motion, and irregular chaotic
motion with and without time-reversal symmetry or Kramers
degeneracy.

We calculate spectra as function of three parameters, that is
deformation as frequency ratios, and spin-orbit and Zeeman strengths.
To investigate the statistics we use the so-called unfolding procedure
to remove the overall energy scale along with unimportant obscuring
average properties. The results exhibit a huge variety of the nearest
neighbor distributions.  Several parameter limits produce simple and
well-known spectra especially obvious when one of the three terms in
the hamiltonian is overwhelmingly dominating. The same
  conclusion about emerging stability holds when the two spin-orbit
  coupling terms are extremely unevenly weighted. All these  cases
are analytically solvable, i.e. either a cylindrical oscillator, a
spin-orbit term or a Zeemann field without the other two.  In
addition, a large deformation is equivalent to a one-dimensional
system with very simple oscillator-like properties for all values of
spin-orbit and Zeeman terms.  These limits are uninteresting for
statistical properties, and intermediate parameter values are
necessary to get sufficiently different couplings.

We first search for structure without Zeeman field. Small
integral frequency ratios are too degenerate to provide irregular
solutions.  For most parameter choices the nearest neighbor
distribution is far from any of the schematic random matrix results.
Systematic resemblance is difficult to find and linear combinations of
Poisson and Wigner distributions are probably the closest similarity.
However, we do find indications of chaotic motion with a
fourth-power level-repulsion term for some irrational frequency ratios
and specific spin-orbit strengths.
Including a finite Zeeman field breaks time-reversal symmetry
and is therefore potentially able to lead to a different chaotic motion
corresponding to a second-power level-repulsion term.  However, we see
a quite different behavior: the histograms for finite values of the
spin-orbit coupling are described very well by an intermediate
Brody distribution.

Our numerical results suggest that a combination of trap deformation,
spin-orbit coupling strength and Zeeman field provides a suitable tool to
manipulate the spectrum.  Some parameters give the Wigner nearest
neighbor distribution which is used as a signature of a quantum chaotic
system.  The spin-orbit coupling appear as a valuable tool to	
investigate regimes of such irregular motion.  This is within
practical reach in present cold atomic gas experiments where a highly
tunable single-particle spin-orbit coupling is already available.
Thus the perspective is that the transition between regular and chaotic
motion can be systematically investigated by varying the spin-orbit
strength for given deformations.  The Zeeman field can also be used to
change the distribution towards another Wigner distribution
corresponding to less level repulsion.

These results can also be looked at from the perspective of a
non-interacting system of $N$ fermions. In this case particles occupy
$N$ energy levels so that the Fermi energy would be equal to the
($N/2$)'th energy level (due to the Kramers degeneracy).  The
dynamics of such systems will be defined by the statistical properties
of the spectrum around the Fermi energy. The requirement is 
tunability of the deformation and subsequent variation of the spin-orbit
strength to move from regular to quantum chaotic systems.  The
experimental capabilities and interest in use of cold atoms
with Rashba-like spin-orbit interactions should make 
fermionic $N$-body systems realizable in the near future.
An interesting question is the role of interactions. 
Within a mean-field Hartree-Fock method using repulsive zero-range interactions, the structure
of the single-particle Hartree-Fock energy levels stays qualitatively the same as the
non-interacting spectrum \cite{marchukov2014}. However, the number of
particles and the strength of interaction are the additional parameters to control the statistical
properties of the spectrum. Thus, a more in-depth analysis of the interacting 
system is an interesting direction for future work.

We thank Jonas Larson for valuable discussions. This work was supported by the 
Danish Council for Independent Research - Natural Sciences.

\end{document}